## Title

- Single-pulse Stimulated Raman Photothermal Microscopy and Direct Visualization of Cholesterol-rich Membrane Domains
- Single-pulse Stimulated Raman Photothermal Microscopy

## Authors

Yifan Zhu[1], Hongli Ni[1], Hongjian He[1], Yueming Li[1], Meng Zhang[1], Ji-Xin Cheng[1,2,3,4,*]

## Affiliations

[1]Department of Electrical and Computer Engineering, Boston University, Boston, MA, 02215, USA.

[2]Department of Chemistry, Boston University, Boston, MA, 02215, USA.

[3]Department of Biomedical Engineering, Boston University, Boston, MA, 02215, USA.

[4]Photonics Center, Boston University, Boston, MA, 02215, USA

[*]Corresponding author: jxcheng@bu.edu

## Abstract

By measuring the thermal effects resulting from stimulated Raman excitations, stimulated Raman photothermal (SRP) microscope offers a new access to spatially and temporally resolved Raman signatures across various sample types. Unlike stimulated Raman scattering (SRS) microscopy, SRP is highly compatible with noisy ultrafast laser sources, allowing the use of high peak power, low repetition rate optical parametric amplifier (OPA) lasers to boost sensitivity and imaging speed. Here, we report a single pulse SRP (spSRP) microscope system in which SRP signals are induced by individual pairs of laser pulses generated by an OPA laser. Extensive pulse chirping is used to maximize SRS excitation rate and to minimize photodamage. The single-pixel limit of detection (LOD) of spSRP on dimethyl sulfoxide (DMSO) reaches 890 μM, which is a ~44-fold improvement over SRS microscope. Versatile applications to fungi, cells, and tissues are demonstrated. Live cell spSRP imaging was carried out at speed of 10 frames per second. Particularly, the spSRP system enables direct visualization of cholesterol-rich domains, co-localized with caveolin immunofluorescence, in the plasma membrane of HeLa cells.

**Teaser:** Single pulse stimulated Raman photothermal imaging provides visual evidence of the long-sought "lipid rafts" in intact cell membranes.

# MAIN TEXT

## Introduction

Coherent Raman scattering (CRS) microscopy (*1*, *2*), including coherent anti-Stokes Raman scattering (CARS) (*3*, *4*) and stimulated Raman scattering (SRS) (*5–9*), has become a cornerstone chemical imaging technique. Owing to its high sensitivity and vibrational contrast, CRS is particularly powerful for visualizing lipid-rich structures (*10–12*) such as intracellular lipid droplets, myelin sheaths, and model membranes by leveraging the strong CH stretch vibration.

Despite these significant advancements, a critical and long-standing challenge has remained largely unresolved: the direct, label-free visualization of membrane domains. These nanoscale domains are hypothesized (*13–15*) to be fundamental organizers of cellular signaling and trafficking, yet their small size, dynamic nature, and subtle chemical differences position them beyond the reach of current CRS imaging. CARS microscopy is hampered by its non-resonant background (*16*), which can mask the weak contrast arising from minor variations in lipid order and cholesterol content within a membrane. While SRS microscopy eliminates this background and offers quantitative contrast, its sensitivity is ultimately constrained by fundamental laser shot noise (*16*). Recent efforts to improve the SRS sensitivity, either provide limited improvement on noise reduction as in quantum SRS (*17–19*) or require sophisticated labeling or plasmonic substrate as in stimulated Raman excited fluorescence (SREF) (*20*, *21*) and plasmon-enhanced SRS (*22*, *23*). Consequently, current CRS microscopy is insufficient to reliably map the subtle chemical signatures of lipid domains in biological membranes.

Stimulated Raman photothermal (SRP) microscopy (*24*, *25*) is a newly developed coherent Raman imaging modality. In SRP measurements, a pair of laser pulse trains is directed to the sample to excite the chemical bond vibrations, while a third probe laser beam measures the thermal lensing effect caused by SRS heating (**Fig. 1A**). By shifting the measurement principle from light absorption (as in SRS) to light refraction (as in SRP), SRP offers three major advantages: large modulation depth, low collection numerical aperture (NA) requirement, and low susceptibility to ultrafast laser noise. These advantages have enabled SRP implementations with distinct features that are vital to biological studies and/or translational applications (*25*).

Current demonstrations of SRP utilize optical parametric oscillator (OPO) systems with high repetition rates (> 40 MHz) and low peak powers (< 10 kW). Meanwhile, optical parametric amplifier (OPA) systems (**Fig. 1B**), noted for their high peak power and broad spectral tunability, are widely used in nonlinear spectroscopy (*26*) and three-photon fluorescence microscopy (*27*). The available femtosecond peak powers of OPA lasers could be ~70-fold higher than that of OPO lasers, which increases the efficiency of nonlinear SRS excitation (**Fig. 1C,D**) by 2 orders of magnitude. Furthermore, the single-pulse configuration reduces the heating period from a burst of pulses (~μs) to a single pulse (~ps), thereby minimizing signal loss from thermal diffusion during heating (**Fig. 1E**). Additionally, the low repetition rate (~1 MHz) of OPA lasers allows sufficient cooling between heating periods without the need for laser modulation or pulse picking, enabling high-speed SRP imaging at rates of up to 1 μs per pixel. Importantly, although OPA lasers are generally too noisy for SRS measurements, they are well-suited for SRP measurements as long as a low-noise probe laser is used. Together, these features make OPA an ideal source for SRP microscopy.

In this work, we present the theoretical modeling and implementation of an OPA-based single pulse SRP (spSRP) microscope, and explore its potential in highly sensitive, high speed chemical imaging. Theoretical simulations indicate that for high peak power excitations, femtosecond pulse durations cause saturation and reduction of excitation efficiency, while extensive pulse chirping to tens of picoseconds substantially enhances the excitation efficiency. In addition to pulse chirping, our spSRP microscope is equipped with a radially segmented balanced detector (*28*) to further boost sensitivity. Versatile applications of spSRP to monitoring fatty acid uptake in cancer cells, heavy water uptake in fungi, brain tissue fingerprinting, and high-speed tracking of lipid droplets in live cancer cells are demonstrated. No photodamage was observed, attributable to extensive chirping of OPA laser pulses. Lastly, we report direct visualization of membrane caveolae enabled by the highly sensitive spSRP microscope.

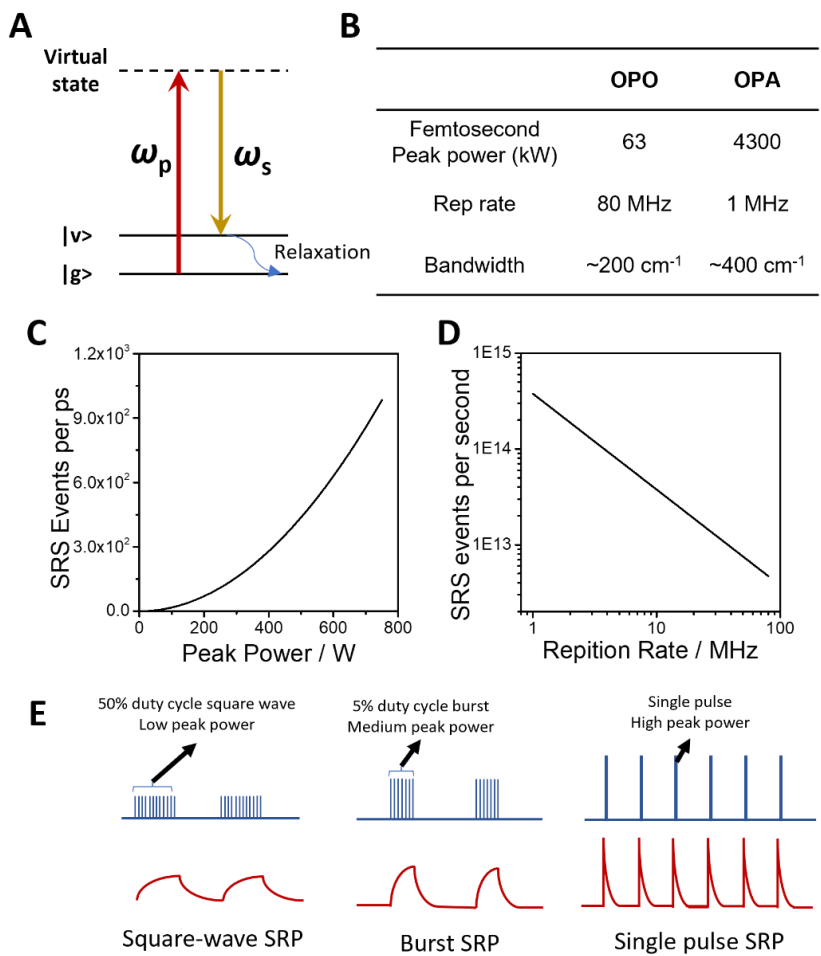

**Figure 1. OPA lasers boost SRS excitation efficiency. A**. SRP energy diagram. The pump and Stokes pulses excite a molecule from ground to excited state. Subsequent vibrational relaxation generates heat. |g> and |v> represent ground and vibrationally excited states. **B**. Key parameters comparison between OPO and OPA lasers. **C**. Nonlinear power dependence of SRS excitation rate. The x-axis denotes the peak powers of both pump and Stokes beams. **D**. Reciprocal relationship between laser repetition rates and relative SRS excitation rates. **E**. Schematic of square-wave SRP, burst-mode SRP, and single pulse SRP. Parameters for **C**: Sample: 14 M (pure) DMSO; Raman peak: C-H asymmetric stretching at 2913 $cm^{-1}$ with cross section of 1 GM; repetition rate: 80 MHz; pulse duration: 5 ps. Focusing NA: 1.2. Parameters for **D**: Sample: 14 M (pure) DMSO; Raman peak: C-H asymmetric stretching at 2913 $cm^{-1}$ with cross section of 1 GM; average power: 30 mW for both pump and Stokes; pulse duration: 5 ps. Focusing NA: 1.2.

## Results

### *Optimal pulse chirping to maximize SRS rate under high peak power condition*

SRS can be modeled as a two-photon vibrational excitation process with absolute cross sections in the unit of Goeppert-Mayer (GM, $10^{-50}$ $cm^4 \cdot s \cdot photon^{-1}$) (*29*). Since the SRS process deposits energy to molecules in the focus, the SRS rate is central to the sensitivity of SRP microscopy, and can be most effectively maximized by increasing the peak laser power (**Fig. 1C**) (*24*). Lasers operating at low repetition rates are particularly well-suited for this purpose, allowing higher peak intensities for a given average power, thereby enhancing the yield of nonlinear signal (**Fig. 1D**).

However, a critical consideration is the onset of transition saturation, a nonlinear phenomenon that occurs under high-intensity excitation. Transition saturation originates from the SRS excitation-induced depletion of the ground vibrational state population, as molecules are promoted to the vibrationally excited states (**Fig. 2A**).

To quantitatively estimate saturation in SRS, a kinetics model was developed with its details provided in Supplementary material. Briefly, the SRS excitation and decay are modeled as first-order chemical reactions, where the rate of excitation is calculated based on the absolute cross section of SRS (*30, 31*), and the rate of decay is obtained from the vibrational state lifetime (*32*).

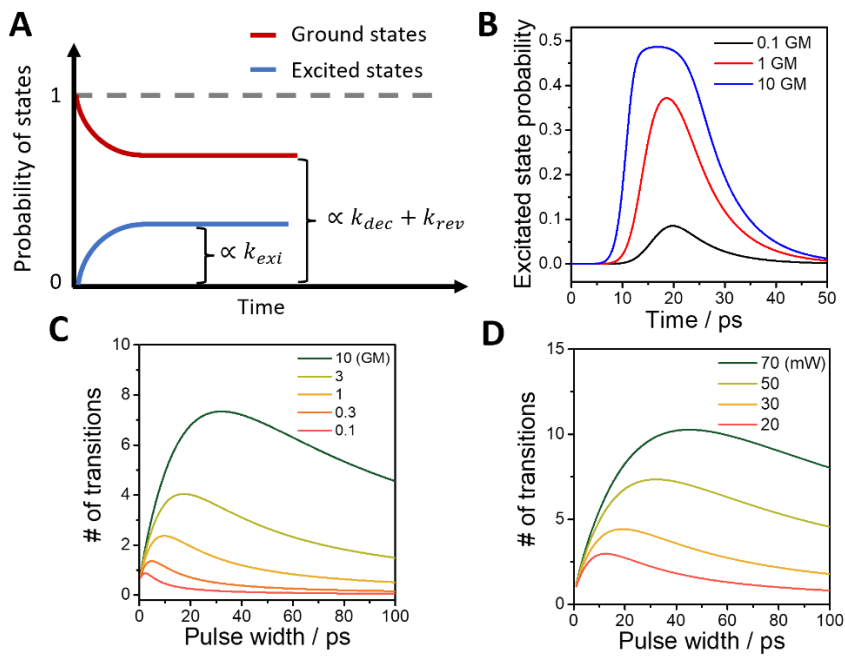

**Figure 2. SRS transition rate as a function of excitation pulse duration. A**. SRS from state population perspective. $k_{dec}$, $k_{rev}$ and $k_{exi}$ are rate constants of spontaneous SRS decay, reverse SRS, and SRS excitation, respectively. A square wave envelope starts at $t = 0$ is assumed for excitation pulses to simplify the diagram. **B**. SRS Saturation as a function of time with a Gaussian excitation envelope. **C**. Number of heat-generating SRS transitions as a function of pulse duration and transition cross sections. **D**. Number of heat-generating SRS transitions per pulse, as a function of pulse duration and average laser powers. Parameters for **B**: Average power: 30 mW for both pump and Stokes; pulse duration: 10 ps; vibrational decay lifetime: 5 ps; repetition rate: 1 MHz. Focusing NA: 1.2. Parameters for **C**: Average power: 30 mW for both pump and Stokes; repetition rate: 1 MHz. Focusing NA: 1.2; vibrational decay lifetime: 1 ps. Parameters for **D**: SRS cross section: 10 GM; repetition rate: 1 MHz. Focusing NA: 1.2; vibrational decay lifetime: 1 ps.

We first illustrate the principle of concept of this model with a simply constant SRS fields that start at $t = 0$ (**Fig. 2A**). Under such a condition, the system will eventually ($t > t_0$) reach equilibrium, then the vibrational state population could be found as:

$$[v](t > t_0) = \frac{k_{exi}}{k_{rev} + k_{exi} + k_{dec}} \cdot [g](0) \quad (1)$$

Where $k_{exi}$, $k_{rev}$ and $k_{dec}$ are the kinetic constant of SRS excitation, reverse SRS and spontaneous decay, respectively. $[v](t)$ and $[g](t)$ are the time-dependent population of vibrationally excited state and ground state, respectively. With weak excitation ($k_{exi} \ll k_{dec}$), which is usually the case for low peak power excitations as conventionally used in OPO SRS, the equilibrium vibrational state population is proportional to the kinetic constant of SRS excitation. However, when $k_{exi}$ is sufficiently large as under high peak power excitation, such proportionality fails and $[v](t > t_0)$ gradually converges to $[g](0)$. This is because the strong excitation depletes the ground state, so that the number of molecules effectively decreases. Such a phenomenon could also be found under temporally Gaussian excitations. As illustrated in **Fig. 2B**, strong excitations with 60 kW, 5 ps pulses for both pump and Stokes, could clearly saturate transitions with high (~1 GM) cross sections, such as C-H asymmetric stretching modes (*30*).

The interplay between nonlinear SRS response, saturation, and photodamage establishes an optimal peak power for efficient excitation under given condition. Consequently, appropriate chirping is necessary to maximize the delivery of laser energies for SRS excitation. Notably, reverse SRS process de-excites the vibrationally excited state without generating heat (*30*). Therefore, the rate of spontaneous decay ($r_{dec}$) is integrated to calculate the total heat-generating transitions (Eq. S12). With varying the pulse durations, we simulated the number of heat-generating SRS transitions on a single chemical bond, as a function of pulse duration as shown in **Fig. 2C,D**. With conserved average power and decay lifetime (**Fig. 2C**), femtosecond excitation does not provide optimal excitation efficiency, since the peak power is too high. On the other end of the spectrum, excessive pulse chirping also reduces the excitation efficiency because the SRS excitation rate scales with the peak power in a quadratic manner. Instead, an optimal point at ~20 ps pulse duration could be found for 3 GM SRS cross section, and larger cross section yields longer chirping for optimal excitation. Additional simulations with conserved cross sections but varying average powers (**Fig. 2D**) revealed that higher average power also yields longer chirping, and ~20 ps remains a relative optimum. These results guide the chirping design in our spSRP microscope.

### *Single pulse SRP microscope with balanced detection*

Based on the above calculations, we have developed an spSRP microscope (**Fig. 3A**) with optimal pulse chirping. Briefly, a seed laser at 1045 nm and a NOPA laser are used for SRS excitation. The synchronized pump and Stokes pulse trains are spatial-temporally aligned and chirped to vibrationally excite the sample molecules. A 520 nm continuous wave probe beam is collinearly aligned with the SRS beams to probe the thermal lensing effect. A telescope adjusts the collimation of the probe beam to introduce an axial offset between the probe laser focus and the SRS focus. The three beams are sent to a laser scanning microscope to interact with the sample. Light is collected with a high NA oil condenser. The outcoming laser is radially segmented into two parts, i.e. the inner core and the outer ring, with a 12.7 mm elliptical mirror. The light intensities of two parts of light are detected with two identical detector arms, digitized simultaneously, and subtracted to obtain the balance-detected signal.

Notably, intensive chirping is introduced based on the simulation of saturation to maximize the excitation efficiency. With 16 travels through 15 cm SF11 glass material on common path, and 7 additional travels on Stokes alone, the pulse duration of Stokes reaches 4.2 ps, and >30 ps of pump pulses by estimation (**Fig. S1**). Such a long chirping reduces the peak power of the SRS excitation lasers, which helps efficiently deliver energy to generate more SRS transitions. Meanwhile, it reduces the photodamage brought by the intense OPA lasers.

The ball-lens model that we proposed previously (*24*) predicts that the inner part and outer part of the probe beam carries opposite modulations induced by thermal lensing effect, as shown in **Fig. 3B**. Therefore, we apply the method proposed by Jun and coworkers (*28*) to perform radially segmented detection to increase the signal by two times while suppressing the probe lase noise. As shown in **Fig. S2**, the condenser was aligned to divergently send out the probe beam to a relay lens at 2 focal lengths from the pupil plane. The relay lens could conjugate the pupil plane of the condenser to another 2 focal lengths away, where the probe beam was segmented. The recorded temporal dynamics measured from 5 mM DMSO-d6 dissolved in DMSO (**Fig. 3C**) of the two segmented portions of probe beam carried common mode laser intensity noise from the probe laser, while the signal modulations were in opposite signs. Following subtraction between the two canceled the common mode noise, meanwhile doubled the signal intensity. It was also clearly shown in frequency domain, where a ~7 dB reduction of noise floor at 1 MHz was achieved (**Fig. 3D**). The noise reduction was >15 dB at lower frequency, which helped improve image quality. Notably at high frequency (>6 MHz), balanced detection elevated the noise floor. This is attributed to the elevated shot noise introduced by increased photon number received by two detectors. Complete frequency spectra including detector baseline and representative signal spectrum are shown in **Fig. S3**.

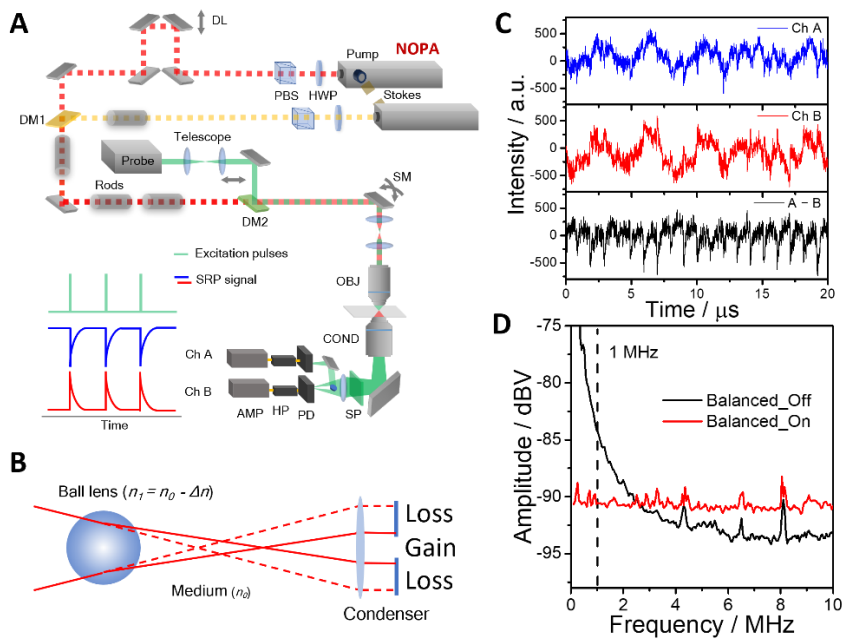

**Figure 3. Detection mechanism and scheme of an spSRP microscope. A**. Scheme of spSRP microscope. DL: delay line. PBS: polarization beam splitter. HWP: half-waveplate. DM: dichroic mirror. SM: scanning mirror. OBJ: objective. COND: condenser. SP: spectral filter. PD: photodiode. HP: high-pass filter. AMP: amplifier. **B**. Ball lens model suggests gain and loss of probe intensities at different detection region. **C**. Time domain SRP signal of 5 mM DMSO-d6 dissolved in DMSO, from each detection channel and their subtraction. **D**. Frequency domain comparison between the probe laser noise (Balanced_Off), probe laser noise after noise cancellation (Balanced_On), as well as a weak SRP signal (140 mM, or 1% DMSO dissolved in DMSO-d6).

*Performance characterization*

The single-pixel limit of detection (LOD) measurement was carried out for DMSO, focusing on the 2913 cm$^{-1}$ mode. To keep the thermal and optical properties constant throughout the measurement, deuterated DMSO (DMSO-d6) was used as the solvent to dilute DMSO. As shown in **Fig. 4A,B**, the SRP spectrum was clean and smooth with a high concentration DMSO sample, and the signal was observable at a concentration as low as 1.7 mM. We calculated the LOD as 890 µM using LOD = 3σ/k, where σ is the standard deviation of the baseline and k is the slope of the intensity-concentration linear calibration curve. In comparison, the LOD by an OPO-based SRP was found to be 2.3 mM (*24*). Thus, an spSRP measurement offers a ~ 2.5-fold improvement. The LODs for C-D bond were measured in DMSO media using DMSO-d6 (**Fig. 4C,D**). Likewise, spSRP showed superior sensitivity to OPO-based SRP, with ~ 1.7-fold LOD increase from 8.4 mM to 5.2 mM. The measured DMSO-d6 spectrum also suggested a high spectral resolution (**Fig. S4**), where the full-width-half-maximum (FWHM) of the 2125 cm$^{-1}$ peak reached 9.9 cm$^{-1}$, close to the Raman spectrum obtained from a spectrometer with 2 cm$^{-1}$ spectral resolution (*33*).

The spectral fidelity and resolution of the system were characterized by imaging polymer nanoparticles, with mixture of 500 nm and 200 nm poly(methyl methacrylate) (PMMA) particles as testbeds (**Fig. 4E-G**). Particles with both sizes could be well-resolved under spSRP, with a signal-to-noise ratio (SNR) of ~120 for 500 nm particles and ~17 for 200 nm particles (**Fig. 4E**). The PMMA spectrum peaked at 2950 cm$^{-1}$ could be recorded with high fidelity, regardless of particle sizes (**Fig. 4F**). The introduction of a third probe beam at 520 nm helped improve the spatial resolution (**Fig. 4G**). We plotted the intensity profile across 3 of 200 nm PMMA beads, with the Gaussian fitted FWHM found to be ~265 nm. Deconvolution with the shape of the beads generated an FWHM of ~ 194 nm, which was below the theoretical resolution limit of SRS under the same condition (~300 nm, FWHM of Airy disk).

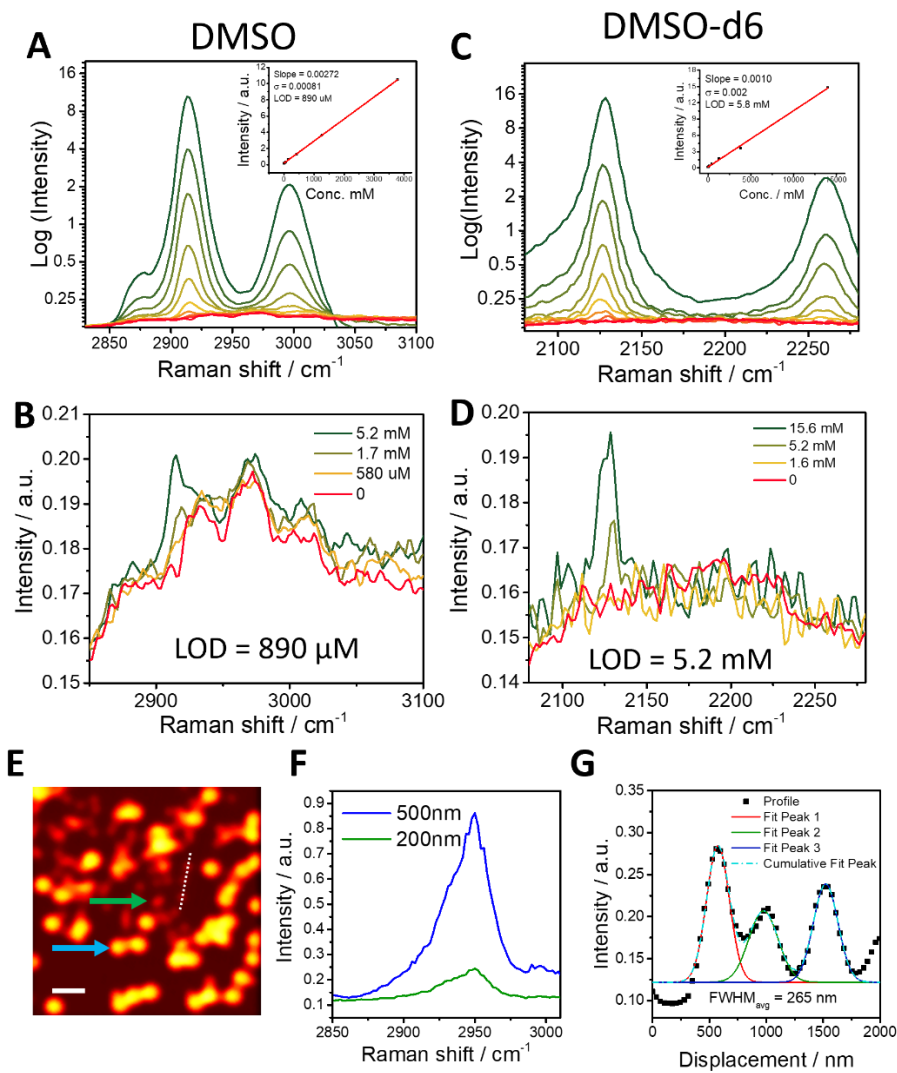

**Figure 4. Performance characterization of spSRP microscope. A-D.** Limit of detection of spSRP microscope. **A**. SRP spectrum (log scale) as a function of DMSO concentration. Insert shows the linear fitting between intensities and concentrations. **B**. Single-pixel SRP spectrum (linear scale) of DMSO at low concentrations. **C**. Single-pixel SRP spectrum (log scale) as a function of DMSO-d6 concentration. Insert shows the linear fitting between intensities and concentrations. **D**. Single-pixel SRP spectrum (linear scale) of DMSO-d6 at low concentrations. Concentration unit: mM. **E-G.** Hyperspectral spSRP imaging of mixture of 500 nm and 200 nm PMMA nanoparticles. **E.** spSRP imaging at 2930 cm$^{-1}$. Scale bar: 1 μm. **F**. Spectra acquired from single 500 nm (blue arrow) or 200 nm PMMA (green arrow) nanoparticles in **E**. **G**. Profile of the white dotted line in **E**. Multiple-peak Gaussian fitting yields average FWHM of ~265 nm.

.

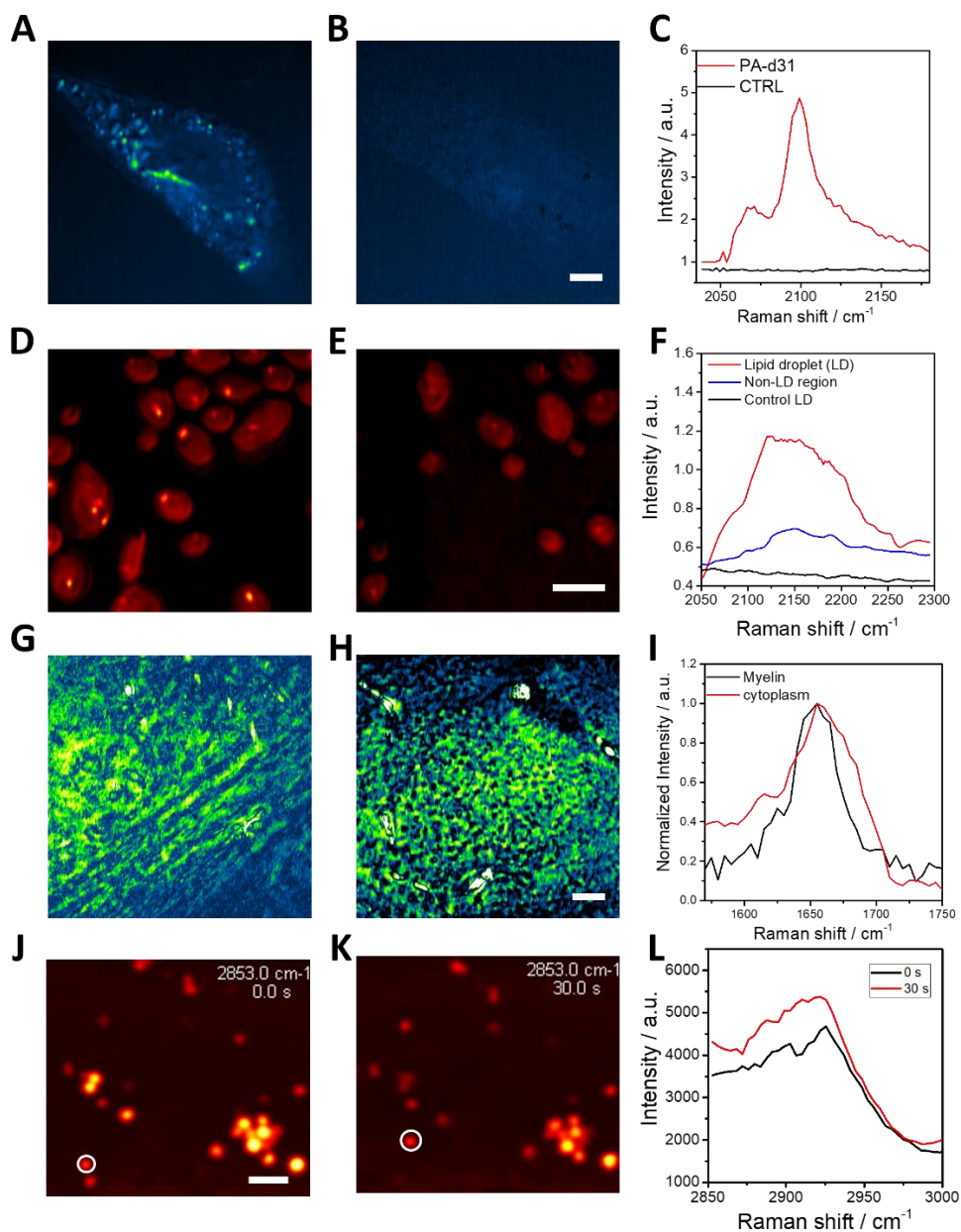

**Figure 5. SRP imaging of bio-samples in native aqueous environments. A-C**. spSRP imaging of PA-d31 cultured SJSA-1 cells. Scale bar: 5 μm. **A**. PA-d31 cultured SJSA-1 cell imaged at 2100 cm$^{-1}$. **B**. Control group with no PA-d31 in culture medium imaged at 2100 cm$^{-1}$. **C**. Spectrum of lipid droplets in PA-d31 cultured cell. No signal was found in the control cell. **D-F**. spSRP imaging of 50% D$_2$O cultured *C. albicans* fungi. Scale bar: 5μm. **D**. 50% D$_2$O cultured *C. albicans* at 2150 cm$^{-1}$. **E**. Control group with no D$_2$O added. **F**. Spectrum comparison between different regions of D$_2$O cultured *C. albicans* as well as control group. **G-I**. spSRP imaging of a mouse brain slice at the fingerprint region. Scale bar: 15 μm. **G,H**. Representative images of different structures of brain (**G**: myelin; **H**: cells). **I**. Spectrum comparison between representative regions. **J-L**: High-speed hyperspectral tracking of lipid droplets in a live HeLa cell. Scale bar: 2 μm. **J** and **K** shows the spSRP image of the cell at 2853 cm$^{-1}$ at t=0 and t=30 s, respectively. **L**: SRP spectrum of the monitored lipid droplet, indicated by the white circle.

## *Biological Applications*

To explore the applicability of spSRP imaging in the silent window, we studied the cellular uptake of deuterated palmitic acid (PA-d31). **Fig. 5A-C** shows the hyperspectral SRP images of SJSA-1 cells incubated with PA-d31. Directly from the SRP signal intensity at the Raman resonance of PA-d31 (2100 cm$^{-1}$), the distribution of PA-d31 formed clusters inside the cell, and the PA-d31 rich regions resembled the morphology of the lipid droplets and ER. The spectral profile (**Fig. 5C**) shows a peak at 2100 cm$^{-1}$, where the C-D stretch vibration resides. In the control sample without PA-d31 treatment, signal showed very weak contrast (**Fig. 5B**), and the corresponding SBR is ~5. This SBR value was much higher than previous result obtained with OPO-based SRP (~1.3) (*24*), which was largely attributed to the elevated peak power and consequently higher ratio between nonlinear and linear photothermal processes.

To demonstrate the feasibility of spSRP imaging for microbiology applications, *C. albicans* metabolism activity was examined by measuring the bio-incorporation of heavy water (**Fig. 5D-F**). **Figure 5D-E** shows the SRP images of *C. albicans* incubated with (**D**) or without (**E**) heavy water. In comparison, the treated group showed much stronger signal than the untreated group, especially at its core lipid droplet. The SRP spectrum (**Fig. 5F**) of the treated group showed a broad peak at C-D stretching window, majorly from the metabolic product of D$_2$O incorporation. The SRP spectrum of control group was flat and featureless, suggesting the non-Raman origin of the signal. Notably, the intensity of the signal in control group decreased as the hyperspectral scanning continues, suggesting a pigment electronic absorption photothermal background. In principle this background could be reduced or eliminated by photobleaching before SRP imaging. During bleaching process, the bond-selective SRP signal is not affected.

We further extend the scope of application to fingerprint region, with mice brain slices as a test bed **Fig. 5G-I**. Representative fields-of-view (FOVs) are selected to demonstrate imaging across diverse brain architectures. **Fig. 5G** clearly revealed myelin sheaths morphology, with a relatively narrow-band C=C signature peak (~1650 cm$^{-1}$) as in the spectrum (**Fig. 5I**, black line), suggesting lipid-dense nature of myelin sheaths. **Fig. 5H**, to the contrary, highlighted densely packed cytoplasmic organelles, whose spectra were predominantly characterized by the relatively broad-band amide I signature (~1650 cm$^{-1}$) (**Fig. 5I**, red line). These results collectively confirm the capability of spSRP for high quality fingerprint imaging in complex tissue environments.

The single-pulse modulation-free measurement scheme further enabled high speed SRP imaging of live cells. As shown in **Fig. 5J-L**, the lipid droplets dynamics inside a live Hela cell were captured. The SRP images were acquired at a speed of 10 frame-per-second (FPS), or 3 s for a 30-frame hyperspectral stack. The white circle highlighted a lipid droplet that was constantly moving during the 36 second data acquisition. With high-speed SRP imaging capability, the spectrum of this moving lipid droplet could be acquired with little motion artifacts (**Fig. 5L**), while it was not possible with previous OPO-based SRP which worked at ~3.5 second per frame (*24*). Representative images (**Fig. S5**) and a complete movie (**Movie S1**) could be found in supplementary information. These data also show that spSRP microscopy with proper pulse chirping induce minimal photodamage to live cells.

*Direct visualization of cholesterol-rich membrane domains*

First proposed in 1997 (*34*), "lipid rafts" are nanoscale, cholesterol and sphingolipid-rich domains (*14, 35*) within cell membranes that act as vital organizing centers for cellular signaling (*36*), trafficking (*37*), and pathogen entry (*13*). Their discovery, through detergent-treated membrane patches, revolutionized our understanding of the plasma membrane, transforming it from a uniform "fluid mosaic" into a dynamically organized platform. However, such "lipid rafts" lack direct visual evidence, owing to their small size, dynamic nature, and lipid constitutes that are difficult to label with fluorophores (*15*). Label-free imaging techniques, on the other hand, lack sensitivity to visualize these nanoscale lipid structures.

The high sensitivity of spSRP enabled the visualization of nanoscale lipid structures that were previously invisible under CARS or SRS. As shown in **Fig. 6**, small, dot-like structures were clearly observed on the membrane of fixed, glycerol-d8 immersed HeLa cell. Here, applying glycerol improves the SRP detection sensitivity by ~3-fold (*24*). Co-localization with anti-caveolin immunofluorescence (**Fig. 6B,C**, **Fig. S6**) suggests that these membrane domains are likely caveolae, which is a crucial structure in cell endocytosis. A signature cholesterol peak (2875 cm$^{-1}$) could be found in the spectrum of these membrane domains (**Fig. 6E**, red line), which is distinct from the neighboring membrane regions (**Fig. 6E**, black line) and internal lipid droplets (**Figure 6E**, blue line) marked in **Fig. 6D**. The spectra of a series of lipid domains (**Fig. 6F**) are displayed in **Figure 6G**, where the cholesterol signature peak could be found in all spectra. Spatial profiling of these membrane domains (**Fig. 6H,I**) and subsequent Gaussian fitting resulted in ~200 nm FWHM, close to the diffraction limit (~200 nm) of the imaging system, indicating that the size of these membrane domains could be much smaller than the diffraction limit. Collectively, by spSRP microscopy, we observed cholesterol-rich membrane structures that colocalizes well with caveolin proteins in the intact plasma membrane environment.

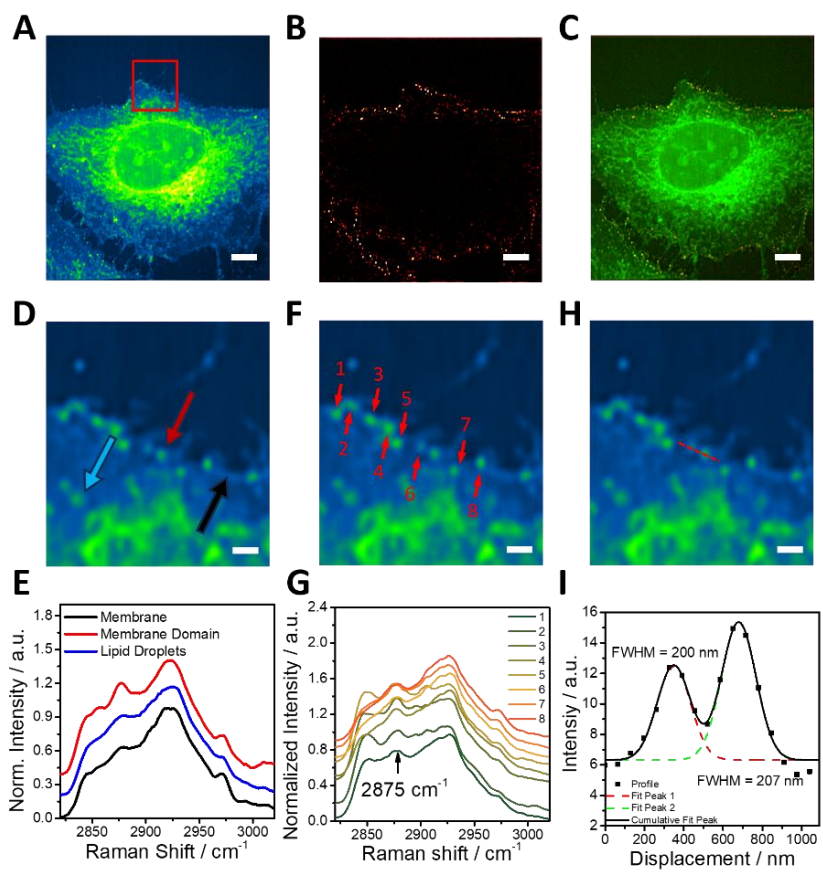

**Figure 6. spSRP enables direct visualization of cholesterol-rich membrane domains. A**. spSRP imaging of a fixed HeLa cell immersed in glycerol-d8. Raman shift: 2875 cm$^{-1}$. Scale bar: 3 μm. **B**. Immunofluorescence stained with anti-caveolin on the same cell. **C**. Co-localization of the spSRP image and immunofluorescence image. **D-I**. Zoom-in analysis of the red-box region in **A**. Scale bar: 600 nm. **D-E**. Spectrum comparison (**E**) between different regions indicated by arrows in **D**. **F-G**. Normalized SRP spectrum (**G**) of different domains on the membrane, indicated by red arrows in **F**. A 0.1 upshift is manually added to the spectrum to improve visibility. **H-I**. Intensity profile and Gaussian fitting (**I**) of the red dashed line in **H**.

## Discussion

We report the theoretical consideration and experimental demonstration of OPA-based spSRP microscopy. We numerically modeled the saturation behavior of SRS under high excitation peak power, which guided the experimental chirping of the pump laser pulse to ~30 ps. We further implemented the radially segmented balanced detection. Together with the high peak power of the OPA laser source, these innovations enabled superior detection sensitivity in comparison to a conventional SRS microscope or an OPO-based SRP microscope. The superior sensitivity of spSRP enabled direct observation of cholesterol rich domains in a plasma member. Below we discuss the advantages, limitations, potential improvements, and future applications of spSRP microscopy.

spSRP is superior in detection sensitivity and imaging speed over OPO-based SRS or SRP microscopies. Thanks to the high laser peak power of an OPA laser, it excites the SRS transitions much more efficiently, which leads to high SRP signals. And the inherent 1 MHz measurement frequency also helps avoid the low frequency noise of the probe laser. Together with the implementation of radially segmented balanced detection which further suppresses the noise floor of the probe laser, we achieved a sensitivity improvement of ~ 2.5-fold comparing to previously reported OPO-based SRP microscope, and ~44-fold comparing to a conventional SRS microscope. Current demonstration of high speed spSRP imaging has reached ~10 frames per second and 3 seconds for a 30-frame hyperspectral stack, which allows spectroscopic tracking of metabolic activity of live cells.

There is space to further improve the spSRP system. Firstly, the difference in spectral bandwidth between the pump and Stokes lasers in the current configuration is the major reason that hinders the further sensitivity improvement. Currently, the bandwidth of the pump laser beam is ~ 300 cm$^{-1}$, while the Stokes bandwidth is only ~ 60 cm$^{-1}$. As a result, after chirping, the pulse duration of the Stokes pulses is only ~1/5 of the pump laser pulses, and the majority part of the pump laser pulse is not temporally overlapped with the Stokes pulses, and the energy of this part only contributes to background and photodamage but not improving the signal. Secondly, current imaging speed is limited by the bandwidth limit of the scanner. By replacing the galvo mirrors, of which the bandwidth is limited to 1 kHz, with a high-speed scanner, ideally video-rate SRP imaging up to 20~30 FPS can be achieved. In such high speed spSRP measurement, one potential issue is low SNR due to the short integration time. However, the SNR could be recovered with denoising algorithms such as SPEND recently developed for removal of non-independent noise in hyperspectral imaging (*38*).

With a matched bandwidth between the pump and Stokes laser pulses, along with a high-performance scanner, we expect to bring ~5-fold sensitivity improvement and ~3-fold speed improvement to the current version of spSRP microscope. With ~200 μM detection sensitivity and ~30 frames per second imaging frame rate, this new technique will open many opportunities in biological applications, such as studying the structures and dynamics of membrane domains, assessment of anti-microbial susceptibilities (*39*), virus-membrane interactions (*9*), and so on.

## Materials and Methods

### *Sample preparation*

SJSA-1 cells: Purchased were from the American Type Culture Collection (ATCC, Cat#: CRL-2098). Cells were cultured in RPMI-1640 culture medium (Gibco), supplemented with 10% (v/v) fetal bovine serum (FBS) and 1% (v/v) penicillin/streptomycin (P/S). All cells were cultured under controlled conditions in a humidified incubator set at 37 °C with a 5% CO2 supply. For the C-D labeled group, the cells were initially cultured until attaching to the glass bottom dishes, then incubated in RPMI-1640 medium containing de-lipid serum for 3 hours. Subsequently, the cells were incubated with 50 μM PA-d31 in the medium for 18 hours. Afterwards, cells were fixed with 10% neutral buffered formalin for 30 min followed by 3 times PBS wash before microscopic imaging.

Mouse brain tissue sample: The mouse was euthanized and perfused transcardially with phosphate-buffered saline (PBS, 1×, PH 7.4, Thermo Fisher Scientific Inc.) solution and 10% formalin, allowing the fixative to circulate throughout the vasculature. After fixation, the brain was extracted and fixed in 10% formalin solution for 24 h to ensure complete fixation. Then, the mouse brain was submerged in a 1× PBS solution and then sliced horizontally into sections with a thickness of 100 µm using an Oscillating Tissue Slicer (OST-4500, Electron Microscopy Sciences).

HeLa cells: For live cell imaging, HeLa cells were cultured with 2 ml of minimum essential medium overnight at 37°C and 5% $CO_2$, then directly sent to imaging. For lipid domain imaging, the HeLa cells were seeded on No.1 glass coverslip with a density of $1 \times 10^5$ ml$^{-1}$ with 2 ml of minimum essential medium overnight at 37°C and 5% $CO_2$. Cells were then fixed with 4% formaldehyde for 15 min. The fixed cells were washed two times with PBS, then the immunofluorescence staining was performed to label the caveolin protein. Afterward the PBS was decanted and 1 drop of glycerol-d8 was added to the cells. Finally, the cells were sandwiched under a piece of cover glass for imaging.

Fungi: Fungal isolates (*C. albicans* SC5314) were initially revived and cultured in a sterile YPD agar plate at 30 °C. Then, fungal isolates were cultured in sterile YPD broth at 30 °C in an orbital shaker (VWR, model 3500I) at a shaking speed of 200 rpm at a tilted angle of 45°. Logarithmic-phase cells were harvested, centrifuged, and then diluted to a concentration at $10^6$ CFU ml$^{-1}$ into $D_2O$-containing RPMI 1640 medium (90% $D_2O$) for metabolic incorporation. To prepare the RPMI 1640 medium, MOPS was used to adjust the pH value of the medium solution to 7.0. The final solution was sterilized by filtering using a 200 nm filter. The cells were centrifuged, washed with fresh1×PBS, and then fixed in 10% formalin solution. The fungal specimen was washed twice using 1×PBS before imaging. Cells were sandwiched between 2 cover glasses (VWR International) for imaging.

*Immunofluorescence*

The fixed HeLa cells were first permeabilized with cold 0.1% Triton X-100 in PBS for 3 min. After being washed with PBS for 3 times, the cells were blocked with 5% Bovine serum albumin (BSA) dissolved in 0.1 M Glycine for 2 hr. Then the anti-caveolin-1 antibody (abcam) was diluted to 1 µg/mL and added to the cell culture for 12 hr. The solution was decanted, and the cells were washed with PBS for 3 times, then incubated with 1 µg/mL secondar antibody (anti-rabbit-igG Alexa 488) for 1 hr in the dark. The stained cells were washed with PBS for 3 times before being sent to imaging.

*spSRP microscope*

A 90/10 beam splitter is inserted into the Spectra-Physics Spirit30 (Newport) pump laser path to extract part of the 1045 nm beam as the Stokes beam in SRS, and the signal output of the NOPA system is used as the pump beam in SRS. The Stokes beam is first chirped with 7 travels through 15 cm SF11 glass rods (Vibronix Inc) to compensate for the imbalanced chirping between two wavelengths. Then these two pulse trains are spatial-temporally aligned and chirped with 16 travels through 15 cm SF11 rods. Electronic delay line is deployed on the Stokes path to control the relative delay hence spectral focusing wavenumber of the measurement. A 520 nm continuous wave probe laser is first adjusted with a telescope for appropriate beam size and collimation, then combined with a dichroic mirror to serve as the probe beam. The three beams are guided to a two-dimensional galvo scanning unit (GVS002; Thorlabs), which is conjugated by a four-focal system to the back aperture of a 60× water objective (NA = 1.2, UPLSAPO60XWIR; Olympus). The NA of the condenser is adjusted to output the probe beam slightly divergently. The pupil plane is then relayed to a small mirror, where the inner and outer parts of the output beam are segmented and measured separately. The detectors are two broadband silicon photodiodes (Hamamatsu) with 50-ohm resistance, a 130 kHz highpass radio frequency filer (Mini-circuits) and a 40 dB low noise amplifier (CMP60116-2; NF Electronics.). A bandpass optical filter (FBH520-40; Thorlabs) is mounted before on the detector to block the SRS beams and allow sole detection of the probe beam. The output signals are digitized by a fast data acquisition card (Alazar card, ATS9462; Alazar Technologies).

**Acknowledgments**
    **Funding:**
National Institutes of Health grant R35GM136223 (JXC)
National Institutes of Health grant R01EB032391 (JXC)
National Institutes of Health grant R01EB035429 (JXC)
    **Author contributions:**
        Conceptualization: Y.Z., J.X.C.
        Methodology: Y.Z., H.N.
        Software: Y.Z.

Investigation: Y.Z.
Visualization: Y.Z.
Resources: H.H., Y.L., M.Z.
Supervision: J.X.C.
Writing—original draft: Y.Z.
Writing—review & editing: Y.Z., J.X.C.

**Competing interests:** J.X.C. declares financial interests in VibroniX Inc. and Photothermal Spectroscopy Corp., which did not fund the study. J.X.C, Y.Z., X.G., H.N., J.Y. are inventors of a patent application (U.S. Provisional No. 63/441,297), submitted by Boston University, that covers stimulated Raman photothermal microscopy. All other authors declare they have no competing interests.

**Data and materials availability:** All data are available upon reasonable request to the corresponding author (jxcheng@bu.edu (J.X.C.))

# Science Advances

Supplementary Materials for

# Single-pulse Stimulated Raman Photothermal Microscopy Enables Direct Visualization of Cholesterol-rich Membrane Domains

Yifan Zhu *et al.*

*Corresponding author. Email: jxcheng@bu.edu

**This PDF file includes:**

    Supplementary Text
    Figs. S1 to S6
    Movie S1 caption

**Other Supplementary Materials for this manuscript include the following:**

    Movies S1

*Science Advances* Manuscript Template Page **18** of **26**

**Supplementary Text**

Modeling of saturation of SRS excitation under high peak power conditions

For molecules with ground vibrational state |g> and a vibrationally excited state |v>, under SRS excitation from pump and Stokes laser pulses with photon flux $\phi(t)$, the rate constant of SRS excitation ($k_{exi}$) at any moment is given by:

$$k_{exi}(t) = \sigma \cdot \phi_p(t)\phi_{St}(t) \quad (S1)$$

Where $\sigma$ is the cross section of the SRS transition, $\phi_p(t)$ and $\phi_{St}(t)$ are the photon flux as a function of time, of pump and Stokes pulses, respectively. The same pump and Stokes laser pulses also induce reverse SRS process with identical rate constant:

$$k_{rev}(t) = \sigma \cdot \phi_p(t)\phi_{St}(t) \quad (S2)$$

The rate constant of spontaneous SRS decay ($k_{dec}$) can be found from the vibrational state lifetime $\tau$:

$$k_{dec} = 0.693/\tau \quad (S3)$$

Note that the rate constant of decay is not time-varying.

Then the rate of SRS excitation ($r_{exi}$), reverse SRS ($r_{rev}$) and decay ($r_{dec}$) could be defined:

$$r_{exi}(t) = k_{exi}(t) \cdot [g](t) \quad (S4)$$
$$r_{rev}(t) = k_{rev} \cdot [v](t) \quad (S5)$$
$$r_{dec}(t) = k_{dec} \cdot [v](t) \quad (S6)$$

Where $[g](t)$ and $[v](t)$ are the number of ground state (|g>) and excited state (|v>) molecules at any given moment, respectively, that satisfy:

$$[v](t) + [g](t) = [g](0) \quad (S7)$$

where $[g](0)$ is the total number of molecules.

For example, under constant excitation fluxes $\phi_p(t)$ and $\phi_{St}(t)$ that starts at $t = 0$, $k_{exi}(t)$ becomes a constant, and the SRS excitation and decay eventually reach equilibrium, defined as:

$$r_{exi}(t) = r_{dec}(t) + r_{rev}(t) \quad (S8)$$

Under such equilibrium:

$$k_{exi} \cdot [g](t) = (k_{dec} + k_{rev}) \cdot [v](t) \quad (S9)$$

The condition in Eq. S9 is illustrated in **Figure 1A**.

The change of excited state population at any given moment could be found as:

$$d[v](t) = (r_{exi}(t) - r_{dec}(t) - r_{rev}(t))dt \quad (S10)$$

Then the population of excited states follows:

$$[v](t) = \int_0^t d[v](t) \quad (S11)$$

And the accumulated number of SRS non-radiative decay events, which is proportional to heat generation, can be found:

$$N_{SRS}(t) = \int_0^t r_{dec}(t)\, dt \quad (S12)$$

By taking in Gaussian photon fluxes that better resemble real cases, with $\phi_0$ as the maximal flux of the pulse, $t_0$ as the temporal center of the pulse, and FWHM as the pulse duration:

$$\phi(t) = \phi_0 \exp\left(-\frac{(t - t_0)^2}{0.361 \cdot \text{FWHM}^2}\right) \quad (S13)$$

Simulation could be carried out by combining Eq. S1~S7 and Eq. S10~S13.

For a single molecule ($[g](0) = 1$), $[v]$ and $[g]$ become the probability of this molecule existing at its excited or ground states, and $N_{SRS}$ becomes the expectation of SRS excitation events that happens on this single molecule. The simulation result for a single molecule is presented in **Figure 2C-D**.

**Fig. S1. Auto-correlation measurement of chirped pump and Stokes pulses.** The maximal measurement range is 50 ps.

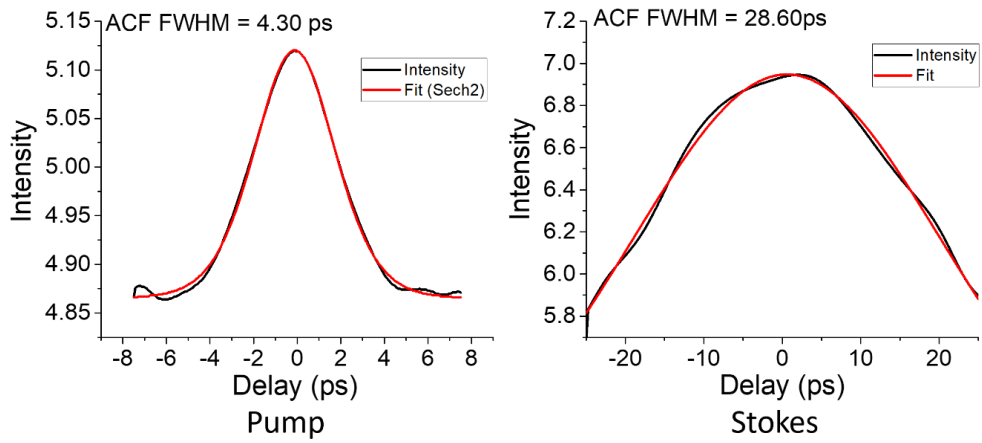

**Fig. S2 Relay relationship between condenser back-pupil and the segmentation mirror for balanced detection.**

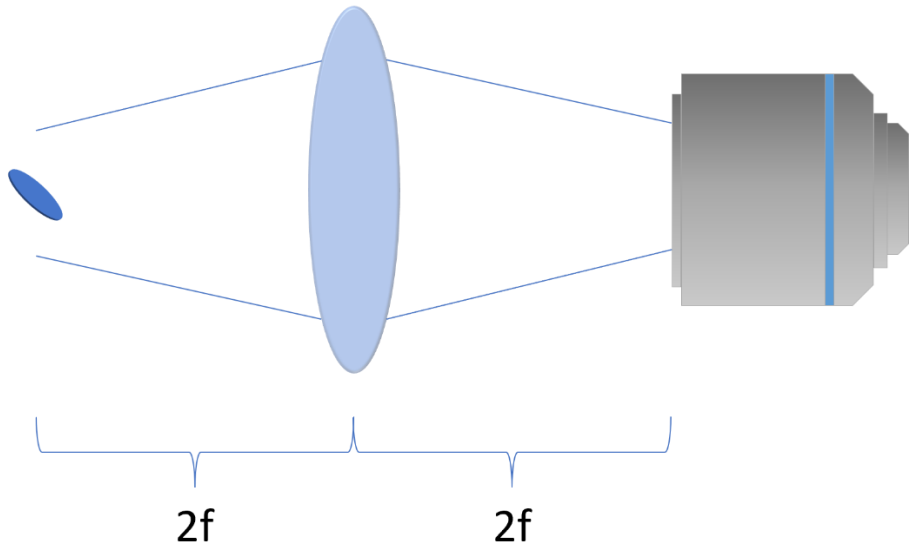

**Fig. S3 Fourier domain analysis of balanced photothermal detection.**

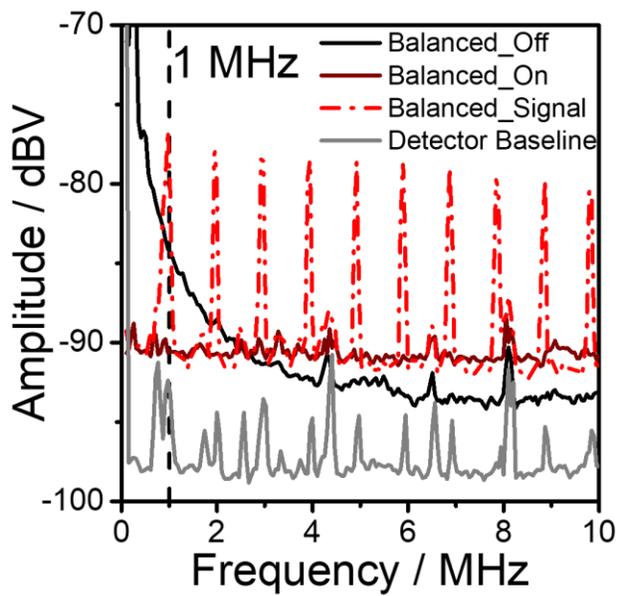

**Fig. S4. Spectral resolution characterization of spSRP microscope, with DMSO-d6 as a testbed.** Lorentz fitting of the 2125 cm$^{-1}$ yields 9.9 cm$^{-1}$ spectral FWHM.

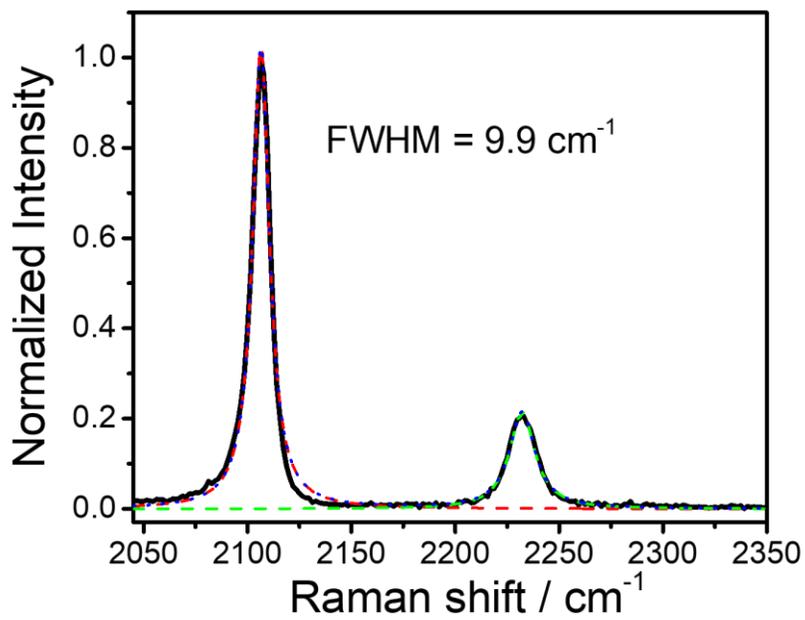

**Fig. S5. Time lapse spSRP images of the live HeLa cell.** Scale bar: 2 μm.

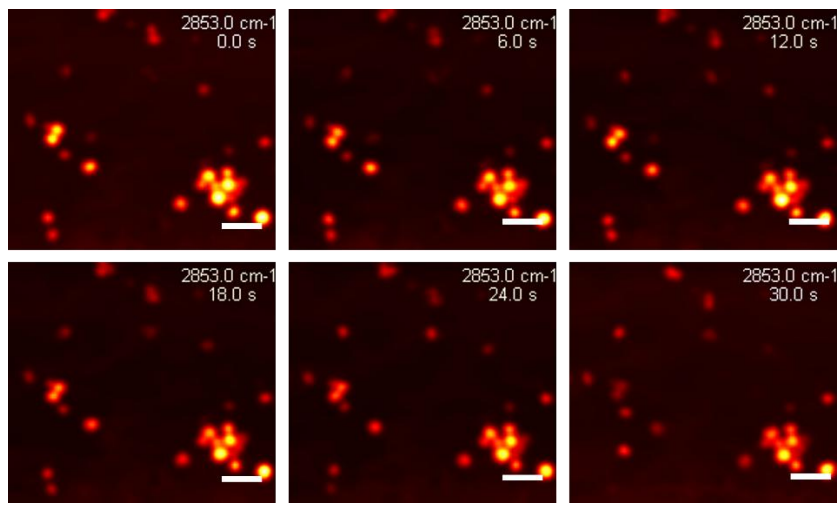

**Fig. S6 Zoom-in of co-localization between the spSRP image of cholesterol rich membrane domains (left) and the immunofluorescence of caveolin (middle).** Scale bar: 600 nm.

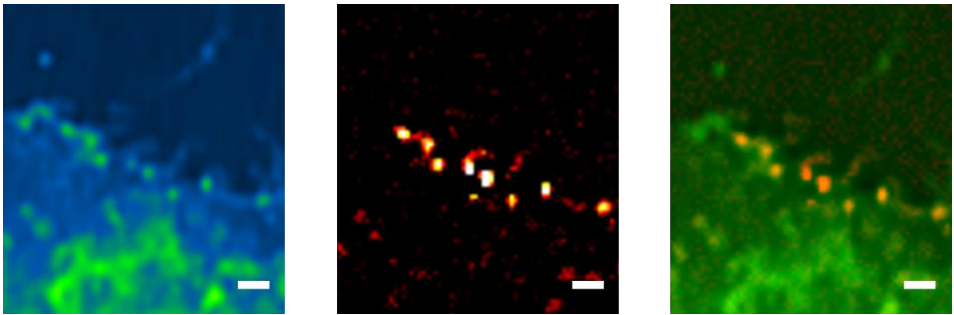

**Movie S1. Hyperspectral spSRP imaging of lipid dynamics inside a living HeLa cell.** Frame rate: 0.1 s/frame, 3 second per hyperspectral stack.